\documentclass{article}

\usepackage{arxiv}

\usepackage[utf8]{inputenc} 
\usepackage[T1]{fontenc}    
\usepackage{hyperref}       
\usepackage{url}            
\usepackage{booktabs}       
\usepackage{amsfonts}       
\usepackage{nicefrac}       
\usepackage{microtype}      
\usepackage{lipsum}
\usepackage{graphicx}
\graphicspath{ {./images/} }

\title{Distributed Kerr-Lens Mode-Locking Based on Spatiotemporal Dissipative Solitons in Multimode Fiber Lasers}

\author{
 Vladimir L. Kalashnikov and
 Stefan Wabnitz\\
 Dipartimento di Ingegneria dell'Informazione\\
 Elettronica e Telecomunicazioni, Sapienza Universit\`a di Roma\\
 via Eudossiana 18, 00184 Rome, Italy \\
Novosibirsk State University, Pirogova 1, Novosibirsk 630090, Russia\\
  \texttt{vladimir.kalashnikov@uniroma1.it, stefan.wabnitz@uniroma1.it}
}

\begin{document}
\maketitle
\begin{abstract}
We introduce a mechanism of stable spatiotemporal soliton formation in a multimode fiber laser. This is based on spatially graded dissipation, leading to distributed Kerr-lens mode-locking. Our analysis involves solutions of a generalized dissipative Gross-Pitaevskii equation. This equation has a broad range of applications in nonlinear physics, including nonlinear optics, spatiotemporal patterns formation, plasma dynamics, and Bose-Einstein condensates. We demonstrate that careful control of dissipative and non-dissipative physical mechanisms results in the self-emergence of stable (2+1)-dimensional dissipative solitons. Achieving such a regime does not require the presence of any additional dissipative nonlinearities, such a mode-locker in a laser, or inelastic scattering in a Bose-Einstein condensate. Our method allows for stable energy (or``mass'') harvesting by coherent localized structures, such as ultrashort laser pulses or Bose-Einstein condensates.
\end{abstract}


\section{Introduction}
The endeavor of multidimensional soliton generation in nonlinear optics (so-called ``light bullets'') and liquid crystals, Bose-Einstein condensates, etc., has a long history \cite{Malomed1,Malomed2,Wabnitz1,Serkin}. Such coherent and strongly localized structures could provide unprecedented energy (or mass) condensation, bridging across micro- and macro-scaled phenomena. The study of multidimensional solitons introduces new branch of ``mesoscopic'' physics, permitting the study of a broad area of nonlinear phenomena far from thermodynamic equilibrium. The main obstacle is that, in contrast with the classical (1+1)-dimensional soliton of the nonlinear Schr\"{o}dinger equation, higher dimensional structures are unstable. Two main approaches have been proposed for the stabilization of a multidimensional soliton, which use i) trapping potentials in a non-dissipative system \cite{Malomed2,Desaix,Wang,Agrawal1}, and ii) nonlinear dissipation \cite{Akhmediev,Malomed3}.

Nonlinear optical systems would furnish an ideal playground in this field, by a ``metaphoric'' (or ``analogous'') modeling, big data, and rare events analysis approach \cite{np}. Specifically, transverse field trapping is an inherent consequence of spatial mode formation in a laser or a passive fiber, where nonlinear effects play a decisive role. In graded-index (GRIN) multimode fibers (MMF), the effect of mode-cleaning, or field self-condensation in the lowest-order spatial modes induced by nonlinear intermodal interaction was recently described \cite{Wabnitz1,Wise1,Wabnitz2,Agrawal2}.

As it was conjectured, a spatially profiled active-ion doping could enhance the beam self-cleaning effect \cite{Beach}. Whereas a mode-locking mechanism provided by the presence of an effective gain, that grows with power (e.g., due to nonlinear polarization rotation), could result in spatiotemporal mode-locking and self-similar pulse evolution in an MMF laser \cite{Wise2,Moser}. The first approach involves using the inherent nonlinearity of a dissipative system for producing a coherent localized, and energy-scalable structure (i.e., a dissipative soliton, DS) \cite{Akhmediev2}. A remarkable breakthrough has been achieved by the development of ultrafast fiber and solid-state waveguide lasers, that allow for avoiding the issues of thermal effects and environmental sensitivity while providing high gains and broad spectral range coverage \cite{Fermann,Mackenzie} and extremely high ultrashort-pulse repetition rates \cite{Kartner}. However, the presence of optical nonlinearities such as self-phase modulation (SPM), four-wave mixing, and stimulated Raman scattering limit ultrashort pulse energy harvesting in fiber lasers. An alternative breakthrough approach was introduced by using solid-state Kerr-lens mode-locked (KLM) oscillators. These sources exploit the effect of loss decrease due to spatial mode squeezing through self-focusing in a nonlinear medium with an aperture \cite{Sibbett}. The evolution of this technology opens the perspective for achieving distributed Kerr-lens mode-locking (DKLM) \cite{Pronin,Wise3}, thus bridging the previously disjointed areas of solid-state and ultrafast fiber photonics, and providing self-spatiotemporal-mode-locking of fiber lasers.

In this Letter, we demonstrate that graded dissipation, provided by  loss/gain transverse profiling in a GRIN fiber, allows for obtaining DKLM in a fiber laser, operating in either anomalous or normal dispersion regimes. The parameters and stability of the resulting spatiotemporal DSs are investigated both analytically and numerically. The problem of the self-emergence (or self-starting) of DSs, and the interdisciplinary outlook for fiber DKLM oscillators are also discussed.

\section{Variational Approximation}

As it was pointed out in \cite{Wabnitz1,Wise4,Wise5}, the Gross-Pitaevskii equation, which is the well-known ``workhorse'' for trapped Bose-Einstein condensate (BEC) modeling \cite{Pitaevskii}, is a well-working approximation for describing pulse propagation in both single and multimode fibers. This equation allows for using the variational approximation (VA) for obtaining a soliton-like solution in a non-dissipative GRIN fiber \cite{Desaix,Wang,Agrawal1,Agrawal2}. The generating Lagrangian $L$ for the Gross-Pitaevskii equation with a parabolic trapping potential can be written as \cite{Agrawal1}:

\begin{eqnarray}\label{eq1}
 L = \frac{i}{2}\left[ {{a^*}{\partial _z}a - a\,{\partial _z}{a^*}} \right] + \frac{1}{2}\left( {{{\left| {{\partial _x}a} \right|}^2} + {{\left| {{\partial _y}a} \right|}^2}} \right) \\ \nonumber + \frac{\delta }{2}{\left| {{\partial _t}a} \right|^2} - \frac{\nu }{2}{\left| a \right|^4} - \frac{s}{2}\left( {{x^2} + {y^2}} \right){\left| a \right|^2},
\end{eqnarray}

\noindent where $a(z,t,x,y)$ is a slowly varying spatiotemporal field profile ($a^*$ corresponds to a complex conjugated value), $z$ is a longitudinal propagation coordinate, normalized to the diffraction length $L_d=\beta_0 w_0^2$, and the transverse spatial coordinates $(x, y)$ are normalized to ${w_0} = 1/\sqrt[4]{{2{k_0}\left| {{n_1}} \right|{\beta _0}}}$. Here $\beta_0=n_0(\omega_0) k_0$ is a propagation constant, $k_0=\omega_0/c$ is a wavenumber, and $n_0(\omega_0)$ is a refractive index on a carrier frequency $\omega_0$. $n_1$ defines a ``curvature'' of the transverse refractive index variation, so that whether $s=+1$ or $s=-1$ corresponds to anti- or guiding GRIN fiber, respectively. The group-velocity and the group-velocity dispersion (GVD) parameters are ${\beta _1} = {\left( {{{d\beta } \mathord{\left/
 {\vphantom {{d\beta } {d\omega }}} \right.
 \kern-\nulldelimiterspace} {d\omega }}} \right)_{\omega  = {\omega _0}}}$ and ${\beta _2} = {\left( {{{{d^2}\beta } \mathord{\left/
 {\vphantom {{{d^2}\beta } {d{\omega ^2}}}} \right.
 \kern-\nulldelimiterspace} {d{\omega ^2}}}} \right)_{\omega  = {\omega _0}}}$, respectively ($\beta=n_0(\omega) \omega/c$). The last parameter defines the local time $t$ normalization to $T_0=\sqrt {\left| {{\beta _2}} \right|{L_d}}$, so that whether $\delta=+1$ or $\delta=-1$ corresponds to the anomalous or normal GVD, respectively. The instantaneous local field intensity $|a|^2$ is normalized to ${k_0}{n_2}{L_d}$ ($n_2$ is a nonlinear refractive index, defining SPM), so that whether $\nu=+1$ or $\nu=-1$ corresponds to a self-focusing or defocusing nonlinearity, respectively (see Table \ref{tab:table1}).

 The dissipative generalization of Eq. (\ref{eq1}) consists in the addition of a ``force" $Q-$term in the Euler-Lagrange equations, in agreement with the Kantarovitch's method (see Supplementary materials and \cite{Kant}):

 \begin{eqnarray}
\label{eq2}
   \frac{{\delta \int\limits_{ - \infty }^\infty  {Ldt} }}{{\delta {\rm{f}}}} - \frac{d}{{dz}}\frac{{\delta \int\limits_{ - \infty }^\infty  {Ldt} }}{{\delta {\rm{f}}}} = 2\Re \int\limits_{ - \infty }^\infty  {Q\frac{{\delta a}}{{\delta {\rm{f}}}}}, \\ \nonumber
 Q =  - i\Lambda a + i\,\tau \,{\partial _{t,t}}a - i\,\kappa \left( {{x^2} + {y^2}} \right)a,
 \end{eqnarray}

\noindent where $\Lambda$ is the difference between loss on the fiber axis and saturated gain. This parameter depends upon the soliton energy $\int_{ - \infty }^\infty  {{{\left| {a\left( {z,x,y,t'} \right)} \right|}^2}dt'}$, and could contribute to the soliton dynamics and stability (see Supplement material and \cite{Beach}). $\tau$ is the spectral dissipation parameter, defined by the inverse of the squared spectral filter (e.g., gain) bandwidth, and $\kappa$ defines the growth of loss along the radial coordinate. The last parameter is determined by graded gain/loss-doping, or by leaking loss in a fiber laser (see Fig. 1). This allows for implementing the general principle of DKLM, that is, the growth of effective gain with intensity, owing to graded loss in a fiber, similarly to the action of a soft or hard aperture in a solid-state KLM laser \footnote{The used notions can vary for different physical models. For instance, the DS energy, phase, and local time coordinate for an optical system correspond to the number of bosons (mass), momentum (wavenumber), and transverse spatial coordinate for a BEC, respectively. Thus, the condensed matter analog of a DS is a formation of a BEC phase.}.

\begin{table}
 \caption{\label{tab:table1}%
The normalization parameters for a GRIN Yb-fiber laser with dispersion compensation, and 10 nm spectral bandwidth. $E_0$ and $P_0$ correspond to the normalization for energy and peak power, respectively. The length of fiber with compensated dispersion is of 3 m. The ``aperture size'' defines the zero-level of net-gain for $\left| \Lambda  \right|=-0.002$, and $\kappa=0.001$.}
  \centering
  \begin{tabular}{lll}
    \cmidrule(r){1-2}
    Name     & Description\\
    \midrule
\textrm{wavelength}&
\textrm{1.06 $\mu$m}\\
$n_0$ & 1.48\\
$n_1$ & 0.02 cm$^{-4}$\\
$n_2$ &$3.5 \times {10^{ - 16}}$ W/cm$^2$\\
$\beta_2$ &30 fs$^2$/cm\\
$w_0$ &83 $\mu$m\\
$L_d$ &6 cm\\
$T_0=\sqrt {\left| {{\beta _2}} \right|{L_d}}$  &230 fs\\
$E_0=w_0^2 T_0/(k_0 L_d n_2)$ &128 nJ\\
$P_0=w_0^2/(k_0 L_d n_2)$ &550 kW\\
$\tau$ &0.4\\
``aperture size'' $d = w_0^2\sqrt {{{\left| \Lambda  \right|} \mathord{\left/
 {\vphantom {{\left| \Lambda  \right|} \kappa }} \right.
 \kern-\nulldelimiterspace} \kappa }}$ &118 $\mu$m
  \end{tabular}
\end{table}

\begin{figure*}
\includegraphics[width=15cm]{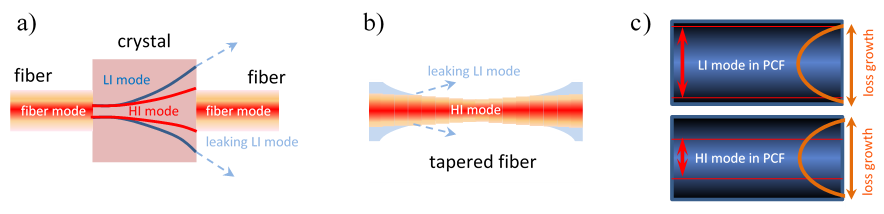}
\caption{\label{fig1}Possible realizations of DKLM in MMF laser: a) disjointed structure with a free beam evolving in a highly nonlinear crystal, leading to highly reduced beam overlap with the subsequent fiber for a low-intensity field (LI mode); b) tapered fiber with reduced leaking mode losses for a high-intensity field (HI mode); c) MMF or photonic-crystal-fiber (PCF) structure with graded losses, providing a loss decrease due to switching to a ``self-focused'' HI mode. The common principle is a self-amplitude modulation by loss decrease for the HI field, in analogy with the KLM principle.}
\end{figure*}

The reduced Lagrangian is calculated using a trial function, corresponding to a soliton-like Gaussian mode:

\begin{equation}
a\left( {z,t,x,y} \right) = \alpha \left( z \right){\mathop{\rm sech}\nolimits} \left( {\frac{t}{{T\left( z \right)}}} \right)\exp \left[ {i\left( {\phi \left( z \right) + \psi \left( z \right){t^2} + \theta \left( z \right)\left( {{x^2} + {y^2}} \right)} \right) - \frac{{{x^2} + {y^2}}}{{2\rho \left( z \right)}}} \right].
 \label{eq3}
\end{equation}

\noindent Here the ${\rm{f}} = (\alpha, T, \phi, \psi, \theta, \rho)-$ parameters describe the $z-$dependent pulse amplitude, duration, phase-delay (${\partial _z}\phi$ could be interpreted as a DS wave-number), chirp, wave-front curvature (spatial chirp), and beam size, respectively. The variation ${{\delta *} \mathord{\left/
 {\vphantom {{\delta *} \delta }} \right.
 \kern-\nulldelimiterspace} \delta }{\rm{f}}$ in Eq. (\ref{eq2}) is performed over these parameters (see Supplemental material for details).

\begin{figure}[b]
\includegraphics[width=10cm]{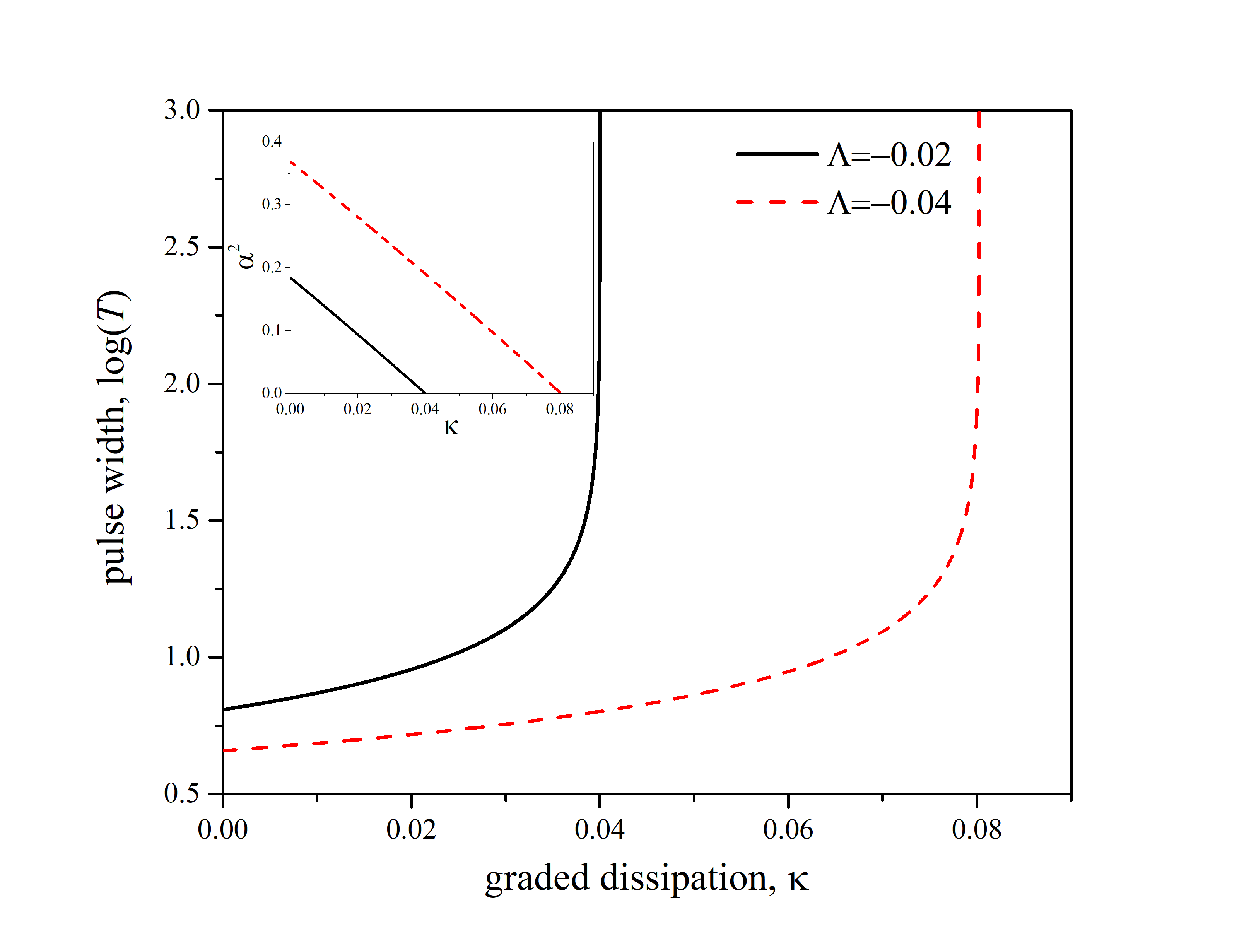}
\caption{\label{fig2} Dependence of the DS temporal width and intensity (inset) on the grading dissipation parameter $\kappa$, for two values of the saturated net-loss $\Lambda$, and $\delta=-1$.}
\end{figure}

The VA demonstrates the existence, above a certain power threshold, of a locally stable non-dissipative soliton. The dimensionless peak power threshold is of $P_0>$5.58 (see Supplemental material) for a guiding GRIN MMF in the anomalous GVD regime ($s=-1$, $\delta=1$) \cite{Agrawal1}. However, such a soliton has a narrow ``attracting basin'', which physically means the impossibility of its ``self-starting'' (``self-emergence'') from an arbitrary initial seed.

Therefore, it is natural to conjecture that such a ``self-emergence'' could exist for a soliton supported by dissipation. The VA-based analysis demonstrates the existence of chirp-free solitons with nonzero wave-front curvature in an anomalous GVD regime ($\delta=1$) for both guiding ($s=-1$) and anti-guiding ($s=1$) refractive index grading. Spatially graded dissipation supports spatial confinement in both cases, under the condition of the absence of spectral dissipation (i.e., $\tau=0$). The DS parameters are:

\begin{eqnarray}
{\alpha ^2} = \frac{{3\left( {{\Lambda ^4} - s{\Lambda ^2} - {\kappa ^2}} \right)}}{{\kappa \,\Lambda \nu }},\\{T^2} = \frac{{2\delta }}{{\nu {\alpha ^2}}}, \nonumber
{\rho ^2} =  - \frac{\Lambda }{\kappa },\,\theta  = \frac{\Lambda }{2}.
 \label{eq4}
\end{eqnarray}

Eqs. (4) demonstrate that the spatial structure of this type of DS is formed by the graded dissipation confinement with an effective aperture size $\chi  = \sqrt {{{\left| \Lambda  \right|} \mathord{\left/
 {\vphantom {{\left| \Lambda  \right|} \kappa }} \right.
 \kern-\nulldelimiterspace} \kappa }}$. The DS duration and intensity are inversely related to each other, as it occurs with nonlinear Schr\"{o}dinger solitons. However, at variance with the latter, the DS intensity is determined by both the refractive and the dissipative guiding properties of the fiber. The ``deconfinement'', which occurs for $|\Lambda|\rightarrow \kappa$ and $s=-1$, means that the DS may also exists when the peak power is reasonably low (depending the net-loss $\Lambda$ variation). This situation is of interest for the self-starting of fiber laser mode-locking.

 The Vakhitov-Kolokolov stability criterion ${{dE} \mathord{\left/
 {\vphantom {{dE} {dq}}} \right.
 \kern-\nulldelimiterspace} {dq}} > 0$ \cite{Malomed1}, where $E = \pi {\alpha ^2}T{\rho ^2}$ is a DS energy, and $q = {\partial _z}\phi  = {{\left( {{\kappa ^2} + 5s{\Lambda ^2} - 5{\Lambda ^4}} \right)} \mathord{\left/
 {\vphantom {{\left( {{\kappa ^2} + 5s{\Lambda ^2} - 5{\Lambda ^4}} \right)} {\left( {4\kappa \Lambda } \right)}}} \right.
 \kern-\nulldelimiterspace} {\left( {4\kappa \Lambda } \right)}}$ is a wave-number, demonstrates the local stability (``attracting basin'') of such a DS for saturated gain parameter values $\Lambda  \in \left\{ {0,\, - \sqrt {{{\sqrt {9 + 20{\kappa ^2}}  - 3} \mathord{\left/
 {\vphantom {{\sqrt {9 + 20{\kappa ^2}}  - 3} {10}}} \right.
 \kern-\nulldelimiterspace} {10}}} } \right\}$. However, this ``attracting basin'' is extremely narrow, with respect to the choice of the initial condition of the field $a$. As a result, such a DS cannot be self-emergent.

Therefore, one may conjecture that a self-emergent spatiotemporal DS should be chirped (i.e., $\psi \ne 0$) due to contribution of spectral dissipation $\tau \ne 0$. The physical solution, in this case, corresponds to:

\begin{eqnarray} \label{sys1}
\nonumber \psi  = \frac{{120\,\tau }}{{{\pi ^2}{T^2}\left( {15\,\delta  + \sqrt {15} \sqrt {15\,{\delta ^2} + 128\,{\tau ^2}} } \right)}}, \\ \nonumber
{\alpha ^2} = \frac{{3 + 3\,s\,{\rho ^4} - 3\,{\kappa ^2}\,{\rho ^8}}}{{\nu \,{\rho ^2}}},\\
 \theta  = {{\kappa \,{\rho ^2}} \mathord{\left/
 {\vphantom {{\kappa \,{\rho ^2}} {2,}}} \right.
 \kern-\nulldelimiterspace} {2,}}
  \\
{T^2} = \frac{{2{\rho ^2}\left( {\delta  + \frac{{80{\tau ^2}\left[ {15\delta \left( {{\pi ^2} - 9} \right) + \left( {{\pi ^2} + 3} \right)\sqrt {225\,{\delta ^2} + 1920\,{\tau ^2}} } \right]}}{{{\pi ^2}{{\left( {15\delta  + \sqrt {225\,{\delta ^2} + 1920\,{\tau ^2}} } \right)}^2}}}} \right)}}{{3 + 3\,s\,{\rho ^4} - 3\,{\kappa ^2}\,{\rho ^8}}}.\nonumber
\end{eqnarray}

The resulting fourth-order polynomial equation has an unique physical solution for the beam-area parameter $\rho^2$ is (see Supplemental material):

\begin{equation}\label{sys1}
\begin{array}{l}
\frac{{\left( {12 + {\pi ^2}} \right)\tau }}{{3{\pi ^2}}} + \Lambda \,{T^2} = \frac{{120\delta \tau }}{{{\pi ^2}\left( {\sqrt {15} \sqrt {15{\delta ^2} + 128{\tau ^2}}  + 15\delta } \right)}} + \\
\frac{{2880{\tau ^3}}}{{{\pi ^2}{{\left( {\sqrt {15} \sqrt {15{\delta ^2} + 128{\tau ^2}}  + 15\delta } \right)}^2}}} - \frac{{\kappa \,{\rho ^2}{T^2}}}{2}.
\end{array}
\end{equation}

The dependencies of the DS temporal width and peak power on the graded dissipation parameter $\kappa$ for the case of normal-GVD are shown in Fig. 2. A DS is positively chirped for both anomalous- and normal-GVD regimes ($\psi>0$), and it has a negative wave-front curvature ($\theta<0$).

\section{Numerical Simulations}

Numerical simulations based on the VA (see insets in Fig. 3) demonstrate the existence of a broad attracting basin of a stable (2+1)-dimensional DS. This means the DS self-emergence from an arbitrary initial Gaussian small signal with amplitude $\alpha_0$ (see Supplemental material). Fig. 3 (curves and scatter points) shows that spectral dissipation enhances the DS stability, and broadens the ``attraction basin,'' i.e., the DKLM capability, which is not possible for $\tau=0$. The growth of the graded dissipation index $\kappa$ reduces the stability region, and increases the DS duration. However, at the same time it also reduces the sensitivity to the initial condition $\alpha_0$, which could mean an enhancement of the DKLM ability. As one can see from Fig. 3 (curves and scatter points), the DS stability regions are broader for the case of normal-GVD.

\begin{figure*}
  \begin{minipage}{.48\textwidth}
    \includegraphics[width=\linewidth]{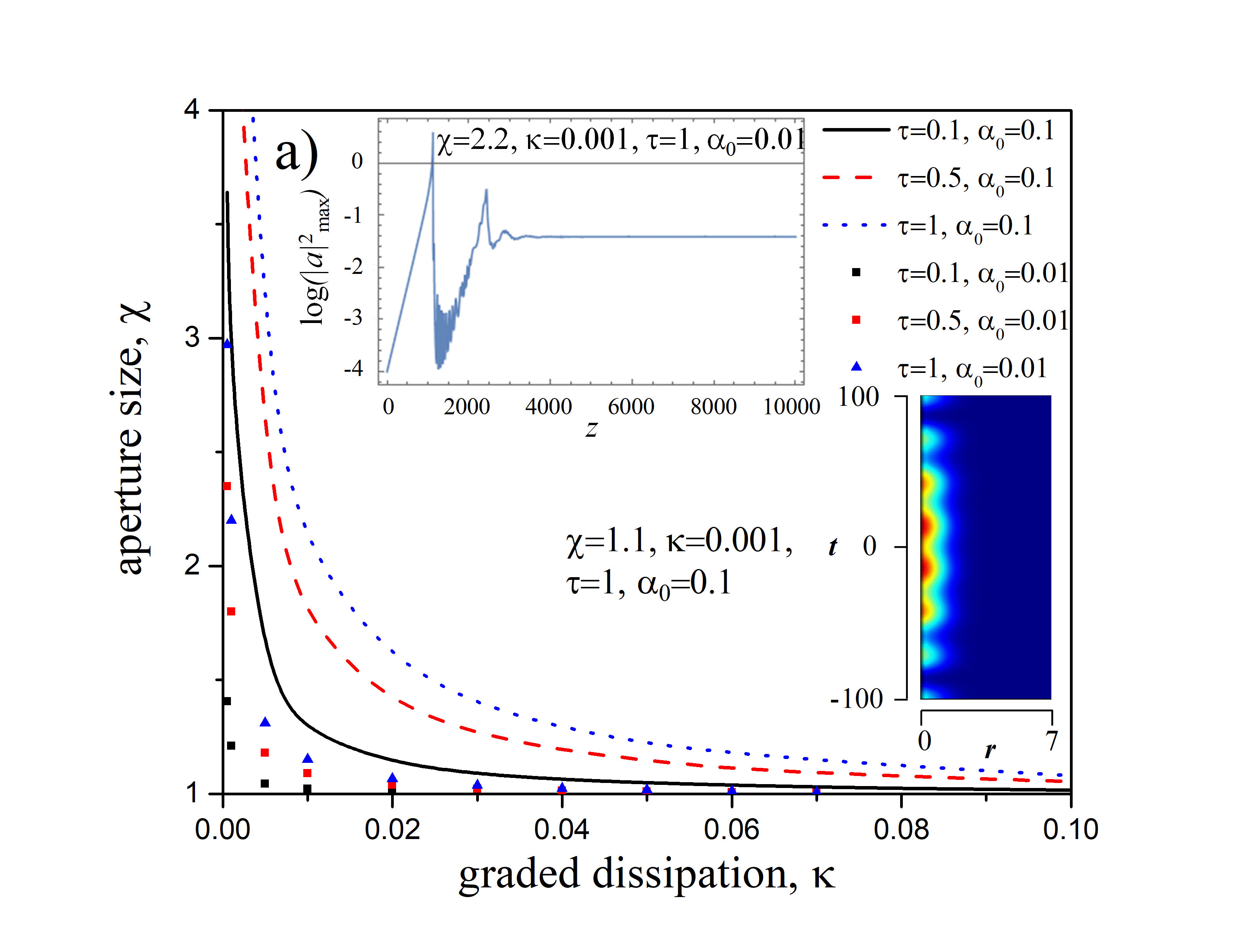}
  \end{minipage} \quad
  \begin{minipage}{.48\textwidth}
   \includegraphics[width=\linewidth]{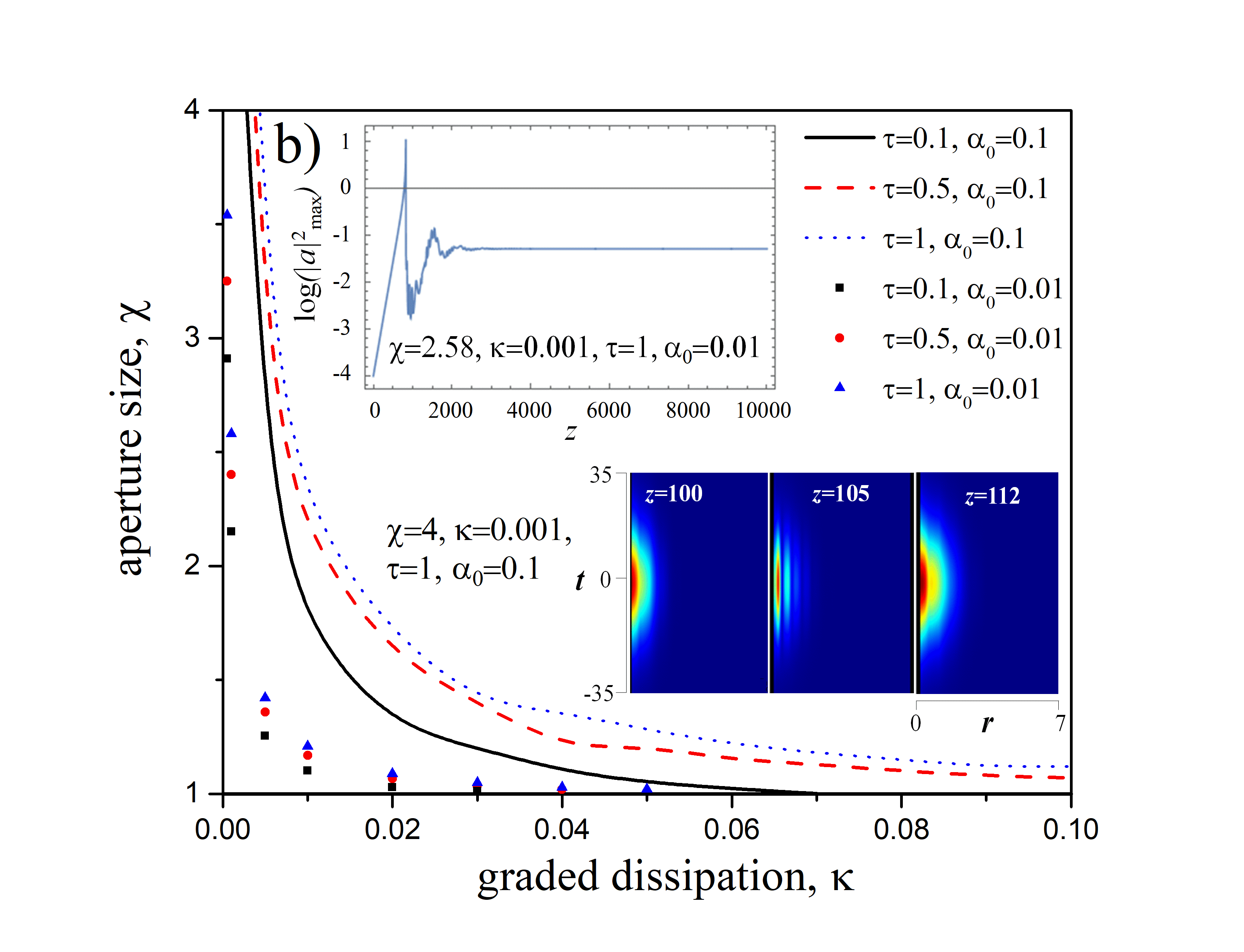}
  \end{minipage}
  \caption{Dependence of the upper boundaries of DS stability in either anomalous- ($\delta=1$, \textit{a}) and normal-GVD ($\delta=-1$, \textit{b}) dispersion regime, respectively, upon the dimensionless ``aperture size'' $\chi=\sqrt{|\Lambda|/\kappa}$, for different spectral dissipation parameters $\tau$. The curve and scatter point plots trace stability boundaries for different initial field amplitude ($\alpha_0$) values. Insets show the DS intensity evolution in the vicinity of the upper stability boundary obtained on the VA based simulations, and contour plots illustrate the DS intensity on the \textit{t} versus $r=\sqrt{x^2+y^2}$ --plane, obtained by full-dimensional Gross-Pitaevskii equation simulations for the shown parameters.}
\label{fig3}
\end{figure*}

Direct numerical simulations of the dissipative Gross-Pitaevskii equation (see Eqs. (\ref{eq1},\ref{eq2})):

\begin{equation} \label{GP}
i\frac{{\partial a}}{{\partial z}} = \frac{1}{2}\left( {\frac{{{\partial ^2}a}}{{\partial {x^2}}} + \frac{{{\partial ^2}a}}{{\partial {y^2}}}} \right) + \frac{\delta }{2}\frac{{{\partial ^2}a}}{{\partial {t^2}}} + \frac{s}{2}\left( {{x^2} + {y^2}} \right)a + \nu {\left| a \right|^2}a - i\Lambda a + i\tau \frac{{{\partial ^2}a}}{{\partial {t^2}}} - i\kappa \left( {{x^2} + {y^2}} \right)a
\end{equation}

\noindent demonstrate the self-starting of DS generation (or DS as a ``global attractor'') for both anomalous- and normal-GVD regimes.

 In addition, full-dimensional Gross-Pitaievskii simulations reveal the contribution of higher-order modes, causing DS oscillations or ``self-imaging'' instabilities \cite{Janner} (inset in Fig. 3, \textit{b}). The main destabilizing scenarios are i) multipulsing in the anomalous-GVD regime (see inset in Fig. 3, \textit{a}), ii) DS collapse in the normal-GVD regime, owing to the presence of a strong Q-switching tendency in the initial stage of mode-locking \cite{Pronin2}, as it is shown in the inset in Fig. 3, \textit{b}, and iii) unlimited temporal spreading of the DS for $\Lambda\rightarrow0$.

\section{Conclusion}

In brief summary, our study demonstrates that exploiting spatially structured dissipative effects may lead to a desirable and feasible breakthrough in mastering energy-scalable and well-controllable spatiotemporal solitons in a fiber self-mode-locked laser. The background approach is to utilize a spatially profiled dissipation (e.g., excitation of leaking radiation, by using waveguide arrays, multicore, or multimode fibers) with the aim of stabilizing the DS, and even providing a robust mechanism of self-starting spatiotemporal mode-locking. In fact, our concept is closely related to the space-time spectral duality \cite{Wise6,Paul} involved in spatiotemporal mode-locking: an initial spatiotemporal multimodal instability interplays with group-delay dispersion and self-phase modulation/self-focusing, from one side, and spectral filtering, from the other side. Such a mechanism of spatiotemporal DS formation can be considered as a path to achieve energy-scalable DKLM in large-mode-area solid-state lasers, in MMF lasers, as well as in photonic lattices.

As an outlook, we anticipate that the nonlinear coupling of spatial modes in either graded-index or photonic-crystal fibers, supported by the presence of graded dissipation, could implement the concept of DKLM in a fiber laser in the regime of multimode self-cleaning. This would provide a means to achieve highly-efficient and stable energy harvesting in an all-fiber laser, without the need of using any additional mode-locking mechanisms. In a broader context, we envisage that photonic devices could provide an efficient tool for metaphorical or analog modeling \cite{np} of strongly localized coherent (or partially coherent) structures, which spontaneously emerge in nonlinear nonequilibrium dissipative systems. In particular, these systems represent a classical analog of the Bose-Einstein condensate in the weakly-dissipative limit.

\section*{Acknowledgments}
This work has received funding from the European Union Horizon 2020 research and innovation program under
the Marie Sk{\l}odowska-Curie grant No. 713694 (MULTIPLY), the ERC Advanced Grant No. 740355 (STEMS), and the Russian Ministry of Science and Education Grant No. 14.Y26.31.0017.  VLK acknowledges the fruitful discussions with Dr. A. Apolonskii inspired the concept of Kerr-lens mode-locking in a fiber laser.

\appendix

\section{Appendixes}
\subsection{\label{sec:ndiss}The variational approach to the non-dissipative Gross-Pitaevskii equation}
Our analytical approach is based on a variational approximation \cite{Agrawal1} to the Gross-Pitaevskii equation. We take into account dissipative factors, which, in particular, are relevant to describing multi-dimensional laser systems. The generating Lagrangian $L$ for the (3+1)-dimensional, nondissipative slowly-varying field amplitude $a(z,x,y,t)$ is
\begin{eqnarray} \label{sL}
 \nonumber  L = \frac{1}{2} \delta ~{\partial _t} a(z,x,y,t) {\partial _t} a^*(z,x,y,t)+\frac{1}{2} \left({\partial _y} a(z,x,y,t) {\partial _y} a^*(z,x,y,t)+{\partial _x} a(z,x,y,t) {\partial _x} a^*(z,x,y,t)\right)+\\  \frac{1}{2} i \left({\partial _z} a(z,x,y,t) {\partial _z} a^*(z,x,y,t)-{\partial _z} a(z,x,y,t) {\partial _z} a^*(z,x,y,t)\right)-\\ \frac{1}{2} s \left(x^2+y^2\right) a(z,x,y,t)
   a^*(z,x,y,t)-\frac{1}{2} \nu  a(z,x,y,t)^2 a^*(z,x,y,t)^2 \nonumber,
\end{eqnarray}

\noindent where $z$ is a propagation coordinate, $t$ is a local time (in a co-moving coordinate system), and $x$, $y$ are transverse coordinates. The parameter definitions and normalizations in Eq. (8) can be found in the main text of the Letter.

The resulting Euler-Lagrange equation is known as the Gross-Pitaevskii equation with a parabolic guiding ``potential''. This equation describes, in particular, beam propagation in a graded-index multimode optical fiber. By using the lowest-mode soliton-like ansatz for a multidimensional soliton (see main text for the definition of the parameters):

\begin{equation}
a_{ansatz}\left( {z,t,x,y} \right) = \alpha \left( z \right){\mathop{\rm sech}\nolimits} \left( {\frac{t}{{T\left( z \right)}}} \right)\exp \left[ {i\left( {\phi \left( z \right) + \psi \left( z \right){t^2} + \theta \left( z \right)\left( {{x^2} + {y^2}} \right)} \right) - \frac{{{x^2} + {y^2}}}{{2\rho \left( z \right)}}} \right].
 \label{sansatz}
\end{equation}

\noindent The subsequent transition to cylindrical coordinates $x = r\cos \chi ,\,\,y = r\cos \chi $ ($r$ is a radial coordinate, $\chi$ is a azimuthal coordinate) allows for obtaining the reduced Lagrangian:

\begin{equation} \label{sL2}
{L_{reduced}} = \int\limits_{ - \infty }^\infty  {\int\limits_0^\infty  {\int\limits_0^{2\pi } {rL\left[ {{a_{ansatz}}} \right]} } } \,dt\,dr\,d\chi.
\end{equation}

The variation of $L_{reduced}$ over the $(\alpha, T, \phi, \psi, \theta, \rho)-$ parameters of the ansatz (9) results in the set of the reduced Euler-Lagrange equations:
\begin{eqnarray} \label{sEL}
\nonumber \frac{2 \delta  \alpha (z) \rho (z)^2}{T(z)}= T(z) \alpha (z)\times \\ \left(\rho (z)^2 \left(6 \rho (z)^2 \left(s+2 \theta '(z)-4 \theta (z)^2\right)+\pi ^2 T(z)^2 \left(\psi '(z)-2 \delta  \psi (z)^2\right) +4
   \nu  \alpha (z)^2+12 \phi '(z)\right)-6\right),\nonumber \\
\nonumber  T(z) \alpha (z) \left(\rho (z)^2 \left(6 \rho (z)^2 \left(s+2 \theta '(z)-4 \theta (z)^2\right)+3 \pi ^2 T(z)^2 \left(\psi '(z)-2 \delta  \psi (z)^2\right)+2 \nu  \alpha (z)^2+12 \phi
   '(z)\right)-6\right)+ \\ \frac{2 \delta  \alpha (z) \rho (z)^2}{T(z)}=0, \\
\nonumber  \alpha (z) \rho (z) \left(\alpha (z) \left(\rho (z) T'(z)+2 T(z) \rho '(z)\right)+2 T(z) \rho (z) \alpha '(z)\right)=0, \\
\nonumber  T(z) \alpha (z) \rho (z) \left(3 \alpha (z) \rho (z) T'(z)+2 T(z) \left(\rho (z) \alpha '(z)+\alpha (z) \left(2 \delta  \rho (z) \psi (z)+\rho '(z)\right)\right)\right)=0, \\
 \nonumber \alpha (z) \rho (z) \left(\alpha (z) \rho (z) T'(z)+2 T(z) \left(\rho (z) \alpha '(z)+2 \alpha (z) \left(\theta (z) \rho (z)+\rho '(z)\right)\right)\right)=0, \\
\nonumber \frac{\alpha (z) \rho (z) \left(-2 \delta +2 T(z)^2 \left(6 \left(\rho (z)^2 \left(s+2 \theta '(z)-4 \theta (z)^2\right)+\phi '(z)\right)+\nu  \alpha (z)^2\right)+\pi ^2 T(z)^4 \left(\psi '(z)-2 \delta
   \psi (z)^2\right)\right)}{T(z)}=0,
\end{eqnarray}
\noindent where prime means a derivative over $z$. System (A4) describes the evolution of the multidimensional soliton parameters (9) with $z$. The steady-state evolution corresponds to $\alpha ' = 0,\,T' = 0,\,\rho ' = 0,\,\psi ' = 0,\,\theta ' = 0,$ and $q=\phi '$ can be considered as a soliton wavenumber.

In the chirp-free case (i.e., both $\psi$ and $\theta$ equal 0, and ${\mathop{\rm sgn}} (\nu ) = {\mathop{\rm sgn}} (\delta )$), there are two soliton solutions:

\begin{eqnarray} \label{sagrawal}
\nonumber {T^2} = \frac{{2\delta }}{{\nu \,{\alpha ^2}}},\,\\
\rho _1^2 = \frac{6}{{\nu \,{\alpha ^2} + \sqrt {{\nu ^2}\,{\alpha ^4} - 36s} }},\\
\nonumber \rho _2^2 = \frac{{\nu \,{\alpha ^2} + \sqrt {{\nu ^2}\,{\alpha ^4} - 36s} }}{{6s}},
\end{eqnarray}

\noindent where the intensity $\alpha^2$ can be treated as a free parameter, per the energy conservation law (the third equation in Eqs.(4)).

\begin{figure*}[b]
\begin{minipage}{.48\textwidth}
\includegraphics[width=8cm]{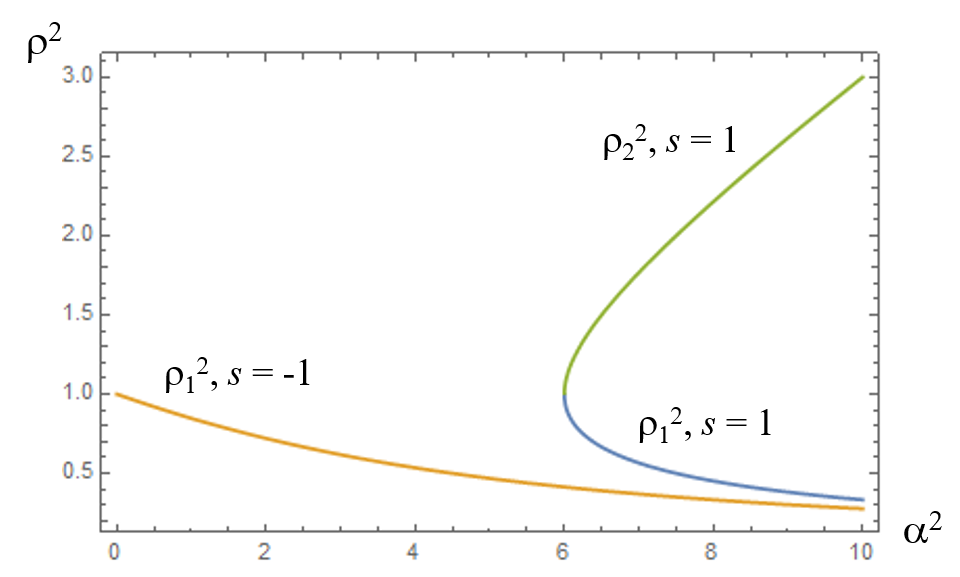}
\caption{\label{fig4} Dependence of the squared soliton beam size on the intensity for three chirp-free solutions (A5); $\nu=1$, $\delta=1$.}
  \end{minipage} \quad
  \begin{minipage}{.48\textwidth}
\includegraphics[width=8cm]{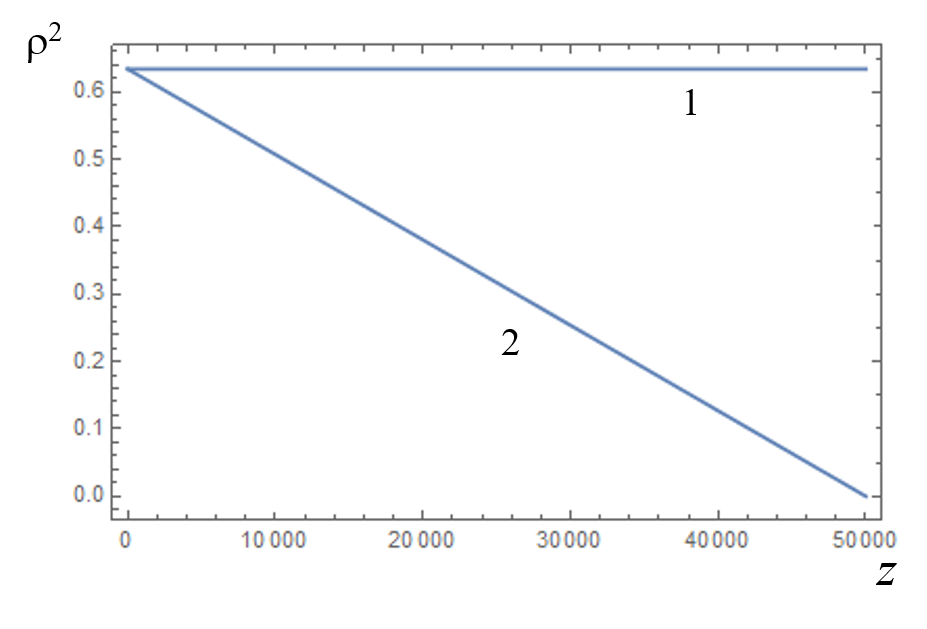}
\caption{\label{fig5} Evolution of the squared beam size $\rho_1^2$ for the exact solution (12) (line 1) and the solution (12) but with $\psi(0)=10^{-5}$ and $\theta(0)=10^{-5}$ in Eqs. (13) (line 2) for $\nu=1$, $\delta=1$, $s=-1$, and $\alpha(0)=2.5$.}
  \end{minipage} \quad
\end{figure*}

In Fig. \ref{fig4} we show the dependence of the squared dimensionless beam size $\rho^2$ (i.e., beam area) on dimensionless peak power $\alpha^2$, for three physical solutions corresponding to either $\rho_1^2$ ($s=-1, 1$) or $\rho_2^2$ ($s=1$).

We will consider the threshold-less (with respect to $\alpha^2$) solution $\rho^2_1$ corresponding to a guiding potential $s=-1$, which is relevant to a fiber laser system.

The Vakhitov-Kolokolov (VK) stability criterion is ${{dE} \mathord{\left/
 {\vphantom {{dE} {dq}}} \right.
 \kern-\nulldelimiterspace} {dq}} > 0$ \cite{Malomed1}, where $E = \frac{6 \sqrt{2} \pi  \alpha  \sqrt{\frac{\delta }{\nu }}}{\alpha ^2 \nu +\sqrt{\alpha ^4 \nu ^2-36 s}}$ is the soliton energy, and $q = {\partial _z}\phi  = \frac{1}{6} \sqrt{\alpha ^4 \nu ^2-36 s}-\frac{\alpha ^2 \nu }{4}$ is a wave-number. The VK criterion demonstrates that there is a stability threshold $\alpha^2>5.58$, which is hardly realistic for a fiber laser.

System (11) results in the following dynamical system for the evolution of the multidimensional soliton parameters:

\begin{eqnarray} \label{sagrawal2}
\nonumber \psi '(z)= \frac{\nu  \alpha (z)^2}{\pi ^2 T(z)^2}-\frac{2 \delta }{\pi ^2 T(z)^4}+2 \delta  \psi (z)^2,\\
\nonumber \theta '(z)= \frac{1}{6} \left(-3 s+\frac{\nu  \alpha (z)^2}{\rho (z)^2}+12 \theta
   (z)^2-\frac{3}{\rho (z)^4}\right),\\
   \phi '(z)= \frac{\delta }{3 T(z)^2}-\frac{7}{12} \nu  \alpha (z)^2+\frac{1}{\rho (z)^2},\\
\nonumber   \alpha '(z)= \alpha (z) (\delta  \psi (z)+2 \theta (z)),\\
\nonumber   \rho '(z)=
   -2 \theta (z) \rho (z),\\
\nonumber   T'(z)= -2 \delta  T(z) \psi (z).
\end{eqnarray}

\noindent Numerical solutions of Eqs.(13) demonstrate, in the presence of temporal- or spatial-chirp weak perturbations, a spatiotemporal collapse-like behavior (see Fig. \ref{fig5}).

\subsection{\label{sec:diss}The variational approach to the dissipative Gross-Pitaevskii equation}

Within the context of our study, the dissipative version of the Gross-Pitaevskii equation can be written in the following form:

\begin{equation} \label{sGP}
i\frac{{\partial a}}{{\partial z}} = \frac{1}{2}\left( {\frac{{{\partial ^2}a}}{{\partial {x^2}}} + \frac{{{\partial ^2}a}}{{\partial {y^2}}}} \right) + \frac{\delta }{2}\frac{{{\partial ^2}a}}{{\partial {t^2}}} + \frac{s}{2}\left( {{x^2} + {y^2}} \right)a + \nu {\left| a \right|^2}a - i\Lambda a + i\tau \frac{{{\partial ^2}a}}{{\partial {t^2}}} - i\kappa \left( {{x^2} + {y^2}} \right)a,
\end{equation}
\noindent where the dissipative terms in the case of a fiber laser can be described as follows: $\Lambda$ is a saturated net-loss on the fiber axis, $\tau$ is the inverse squared bandwidth of the spectral filter, and $\kappa$ is a coefficient of graded dissipation, describing the growth of net loss, starting from the fiber axis and moving toward its periphery. The evaluation of the $\Lambda-$term is not trivial in the general case, because it includes the nonlinear effect of gain saturation, which can be described in the simplest form as:

\begin{equation}\label{sgain}
    - \Lambda  = \frac{{{g_0}}}{{1 + \frac{1}{{{E_s}}}\int_{ - \infty }^\infty  {{{\left| {a\left( {z,x,y,t'} \right)} \right|}^2}dt'} }} - \ell ,
\end{equation}

\noindent where $g_0$ is an unsaturated gain, $E_s$ is a gain saturation energy, and $\ell$ is the loss coefficient on the fiber axis.

Within the context of a weakly dissipative Bose-Einstein condensate, the $\Lambda-$term can be interpreted as a `` velocity of condensation'' from a noncoherent Bose-basin, the $\tau-$term describes an escape velocity from a condensate, growing larger with the Bose-particle kinetic energy, and the $\kappa-$term describes dissipation due to delocalization of the condensate. Then, for the case of $\delta=1$, Eq. (\ref{sGP}) can be reinterpreted as: $z$ corresponds to time, $x, y, t$ correspond to the Euclidian spatial coordinates, and $E$ is the number of particles (mass) of the condensate. The interpretation of the $\delta=-1$ case is less straightforward (nevertheless, see \cite{negative} and the classical analogue such as the Talbot effect \cite{talbot}).

The variational approximation to (\ref{sGP}) consists in the addition of a ``force" $Q-$term in the Euler-Lagrange equations, in agreement with the Kantarovitch's method \cite{Kant}:

 \begin{eqnarray}
\label{seq2}
   \frac{{\delta \int\limits_{ - \infty }^\infty  {Ldt} }}{{\delta {\rm{f}}}} - \frac{d}{{dz}}\frac{{\delta \int\limits_{ - \infty }^\infty  {Ldt} }}{{\delta {\rm{f}}}} = 2\Re \int\limits_{ - \infty }^\infty  {Q\frac{{\delta a}}{{\delta {\rm{f}}}}}, \\ \nonumber
 Q =  - i\Lambda a + i\,\tau \,{\partial _{t,t}}a - i\,\kappa \left( {{x^2} + {y^2}} \right)a,
 \end{eqnarray}

\noindent This results in the following modified equations for the parameters of the ansatz (A2):

\begin{eqnarray} \label{sfull}
\nonumber \psi '(z)= \frac{3 \nu  \alpha (z)^2-4 \left(3+\pi ^2\right) \tau  \psi (z)}{3 \pi ^2 T(z)^2}-\frac{2 \delta }{\pi ^2 T(z)^4}+2 \delta  \psi (z)^2,\\
\nonumber \theta '(z)= \frac{1}{6} \left(-3
   s+\frac{\nu  \alpha (z)^2}{\rho (z)^2}+12 \theta (z)^2-\frac{3}{\rho (z)^4}\right),\\
   \phi '(z)\to \frac{\delta }{3 T(z)^2}-\frac{7}{12} \nu  \alpha (z)^2+\frac{1}{\rho (z)^2}+\frac{1}{9} \left(3+\pi
   ^2\right) \tau  \psi (z),\\
\nonumber   \alpha '(z)\to \frac{1}{15} \alpha (z) \left(3 \pi ^2 \tau  T(z)^2 \psi (z)^2-\frac{5 \left(12+\pi ^2\right) \tau }{\pi ^2 T(z)^2}+15 (-\Lambda +\delta  \psi (z)+2 \theta
   (z))\right),\\
\nonumber   \rho '(z)\to -\rho (z) \left(2 \theta (z)+\kappa  \rho (z)^2\right),\\
\nonumber   T'(z)\to -2 \delta  T(z) \psi (z)-\frac{16}{15} \pi ^2 \tau  T(z)^3 \psi (z)^2+\frac{8 \tau }{\pi ^2 T(z)}.
\end{eqnarray}

A chirp-free ($\psi=0$) solution of (\ref{sfull}) exists only in the absence of spectral dissipation ($\tau=0$), and reads as:

\begin{eqnarray}\label{zero}
\alpha^2= -\frac{3 \left(\kappa ^2-\Lambda ^4+\Lambda ^2 s\right)}{\kappa  \Lambda  \nu },\\
\nonumber T^2= \frac{2 \delta }{\alpha^2 \nu },\, \rho^2= -\frac{\Lambda }{\kappa },\, \theta= \frac{\Lambda }{2}.
\end{eqnarray}

\begin{figure*}[b]
  \begin{minipage}{.48\textwidth}
\includegraphics[width=8cm]{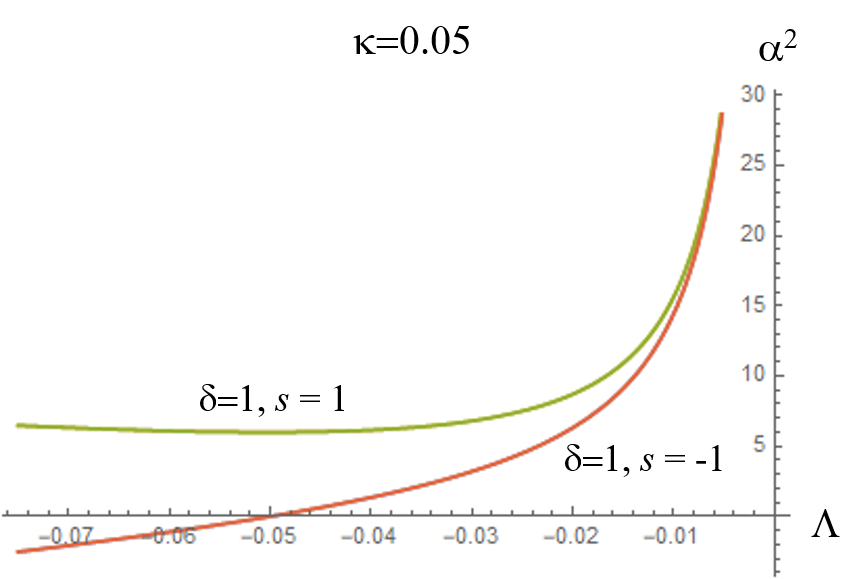}
\caption{\label{fig6} Dependence of the intensity $\alpha^2$ on the saturated net-loss parameter $\Lambda$ for the chirp-free DS.}
  \end{minipage} \quad
  \begin{minipage}{.48\textwidth}
  \includegraphics[width=8cm]{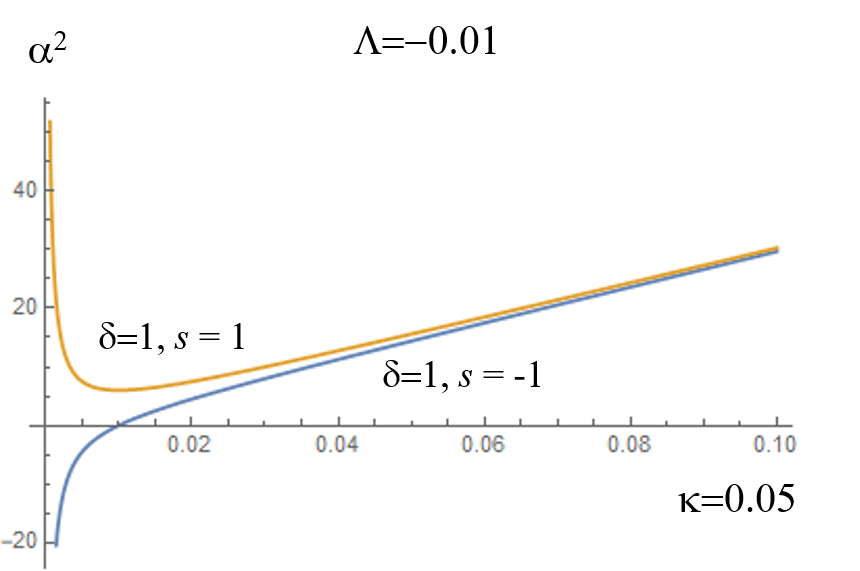}
\caption{\label{fig7} Dependence of the intensity $\alpha^2$ on the graded dissipation parameter $\kappa$ for the chirp-free DS..}
\end{minipage} \quad
\end{figure*}

The only physical configurations are: $\delta=1,\,s=-1$ and $\delta=1,\,s=1$ (i.e., anomalous dispersion combined with guiding/anti-guiding potential, see Figs. \ref{fig6}, \ref{fig7}). A guiding potential provides, for $|\Lambda|\rightarrow\kappa$, relatively low DS peak powers. This is an important property for fiber lasers, as it entails the possibility of achieving the self-starting of passive mode-locking.

\begin{figure*}[b]
  \begin{minipage}{.48\textwidth}
\includegraphics[width=6cm]{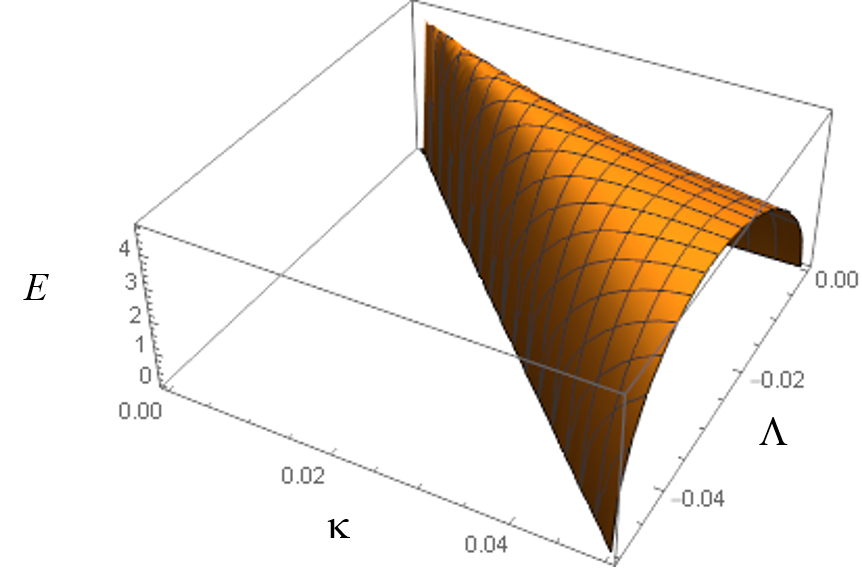}
\caption{\label{fig8} Dependence of the chirp-free DS energy $E$ on the dissipation parameters $\Lambda$ and $\kappa$.}
  \end{minipage} \quad
  \begin{minipage}{.48\textwidth}
  \includegraphics[width=6cm]{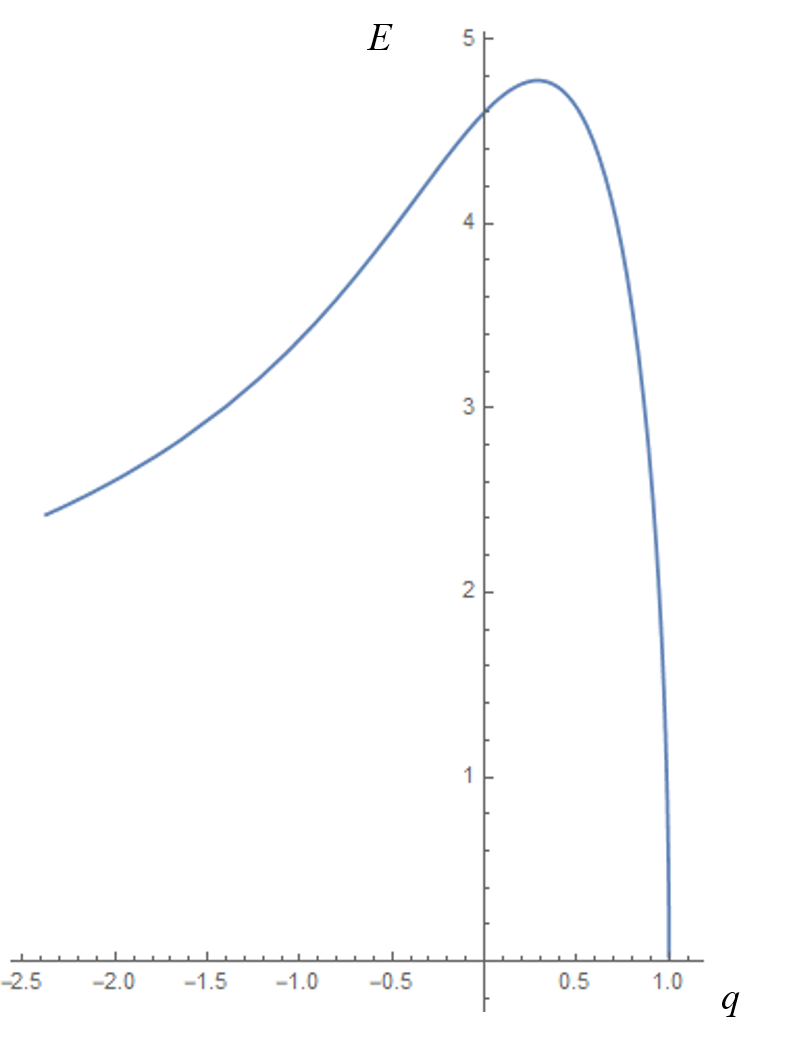}
\caption{\label{fig9} Dependence of the chirp-free DS energy $E$ on the wavenumber $q$.}
  \end{minipage} \quad
\end{figure*}

The nonmonotonous dependence of the DS energy
\begin{equation}\label{sen1}
    E=-\frac{\sqrt{6} \pi  \delta  \Lambda }{\kappa  \nu  \sqrt{-\frac{\delta  \kappa  \Lambda }{\kappa ^2-\Lambda ^4+\Lambda ^2 s}}}
\end{equation}
\noindent on $\Lambda$ (see Fig. \ref{fig8}) suggests its stability, according to the VK stability criterion.

The dependence of the DS energy on the wavenumber

\begin{equation}\label{swn}
    q=\frac{\kappa ^2-5 \Lambda ^4+5 \Lambda ^2 s}{4 \kappa  \Lambda }
\end{equation}

\noindent is shown in Fig. \ref{fig9}.

The VK criterion suggests that DS stability occurs within the parameter region:

\begin{equation}\label{sstab}
   \Lambda  \in \left\{ {0, - \sqrt {\frac{{\sqrt {9 + 20{\kappa ^2}}  - 3}}{{10}}} } \right\}.
\end{equation}

\noindent However, one has to note that the VK stability criterion is not a sufficient condition to guarantee pulse stability. Moreover, a wavenumber cannot be self-consistently formulated in the dissipative case, due to nonlinear nature of $\Lambda$ (see Eq. (15)). Therefore, direct numerical simulations based on the system (17) are required for the case of $\tau=0$.

Our numerical analysis demonstrates the tendency of the beam to collapse, in the presence of weak perturbations of the DS parameters. This conclusion is supported by a linear stability analysis, based on the following stability matrix:
\begin{equation}\label{smatrix}
    \frac{d}{{dz}}\left( {\begin{array}{*{20}{c}}
{{\psi _p}}\\
{{\theta _p}}\\
{{\rho _p}}\\
\begin{array}{l}
{\alpha _p}\\
{T_p}
\end{array}
\end{array}} \right) = \left( {\begin{array}{*{20}{c}}
0&0&0&{\frac{{2\nu {\alpha _s}}}{{{\pi ^2}T_s^2}}}&{\frac{{8\delta  - 2\nu T_s^2\alpha _s^2}}{{{\pi ^2}T_s^5}}}\\
0&{4{\theta _s}}&{\frac{{6 - \nu \alpha _s^2\rho _s^2}}{{3\rho _s^5}}}&{\frac{{\nu {\alpha _s}}}{{3\rho _s^2}}}&0\\
0&{ - 2{\rho _s}}&{ - 2{\theta _s} - 3\kappa \rho _s^2}&0&0\\
{\delta {\alpha _s}}&{2{\alpha _s}}&0&{ - \Lambda  + 2{\theta _s}}&0\\
{ - 2\delta {T_s}}&0&0&0&0
\end{array}} \right)
\end{equation}

\noindent where the $s$ and $p$ subscripts correspond to the steady-state solution and to its perturbation, respectively. Importantly, our analysis shows the crucial contribution of both temporal and spatial chirps in determining beam collapse \footnote{One has to note, that, in the case of the collapse-like behavior, the following assumptions of the considered model become invalid: i) paraxial approximation, ii) lowest-order mode ansatz, and iii) axial symmetry.}.

As a result, one may conjecture that a stable DS soliton has to be necessarily chirped, a condition which requires spectral filtering (i.e., $\tau \neq0$). The system of equations defining the DS parameters reads as:

\begin{eqnarray}\label{ssys}
\nonumber  \alpha^2 \nu +12 \rho^2 \theta^2=\frac{3} \rho^2+3 \rho^2 s, \\
\nonumber  3 \alpha^2 \nu  T^2+2 T^2 \psi \left(3 \pi ^2 \delta  T^2 \psi-2 \left(3+\pi ^2\right) \tau \right)=6 \delta, \\
  \frac{1}{5} \pi ^2 \tau  T^2 \psi^2+\delta  \psi+2 \theta=\Lambda +\frac{\left(12+\pi ^2\right) \tau }{3 \pi ^2 T^2}, \\
\nonumber  2 \theta+\kappa  \rho^2=0, \\
\nonumber  \pi ^2 \psi \left(15 \delta +8 \pi ^2 \tau  T^2 \psi\right)=\frac{60 \tau }{T^2},
\end{eqnarray}

\noindent and the equation for $\phi'(z)$ corresponds to that in Eqs. (17). One obtains the following solutions for the DS intensity $\alpha$ and the wavefront curvature $\theta$:

\begin{equation}\label{ssolution}
\alpha^2= 3\frac{1+\rho^4 s- \kappa ^2 \rho^8}{\nu  \rho^2},\,\theta = -\frac{\kappa  \rho^2}{2}.
\end{equation}

The quadratic equation for the chirp $\psi$:

\begin{equation}\label{schirp0}
-60 \tau +8 \pi ^4 \tau  T^4 \psi^2+15 \pi ^2 \delta  T^2 \psi =0
\end{equation}

\noindent is better to solve by using the Muller's method \cite{muller} to avoid the singularities for $\tau\rightarrow0$.
The obtained physical solution for the chirp-parameter is:

\begin{equation}\label{schirp}
   \psi= \frac{120 \tau }{\sqrt{15} \pi ^2 T^2 \left( \sqrt{\left(15 \delta ^2+128 \tau ^2\right)}+\sqrt{15} \delta  \right)},
\end{equation}

\noindent and the solution for the DS duration is:

\begin{equation}\label{swidth}
    T^2= \frac{2 \rho^2 \left(\frac{80 \tau ^2 \left(\left(3+\pi ^2\right) \sqrt{225 \delta ^2+1920 \tau ^2}+15 \left(\pi ^2-9\right) \delta \right)}{\pi ^2 \left(\sqrt{225 \delta ^2+1920
   \tau ^2}+15 \delta \right)^2}+\delta \right)}{-3 \kappa ^2 \rho^8+3 \rho^4 s+3}.
\end{equation}

\begin{figure*}[b]
  \begin{minipage}{.48\textwidth}
\includegraphics[width=8cm]{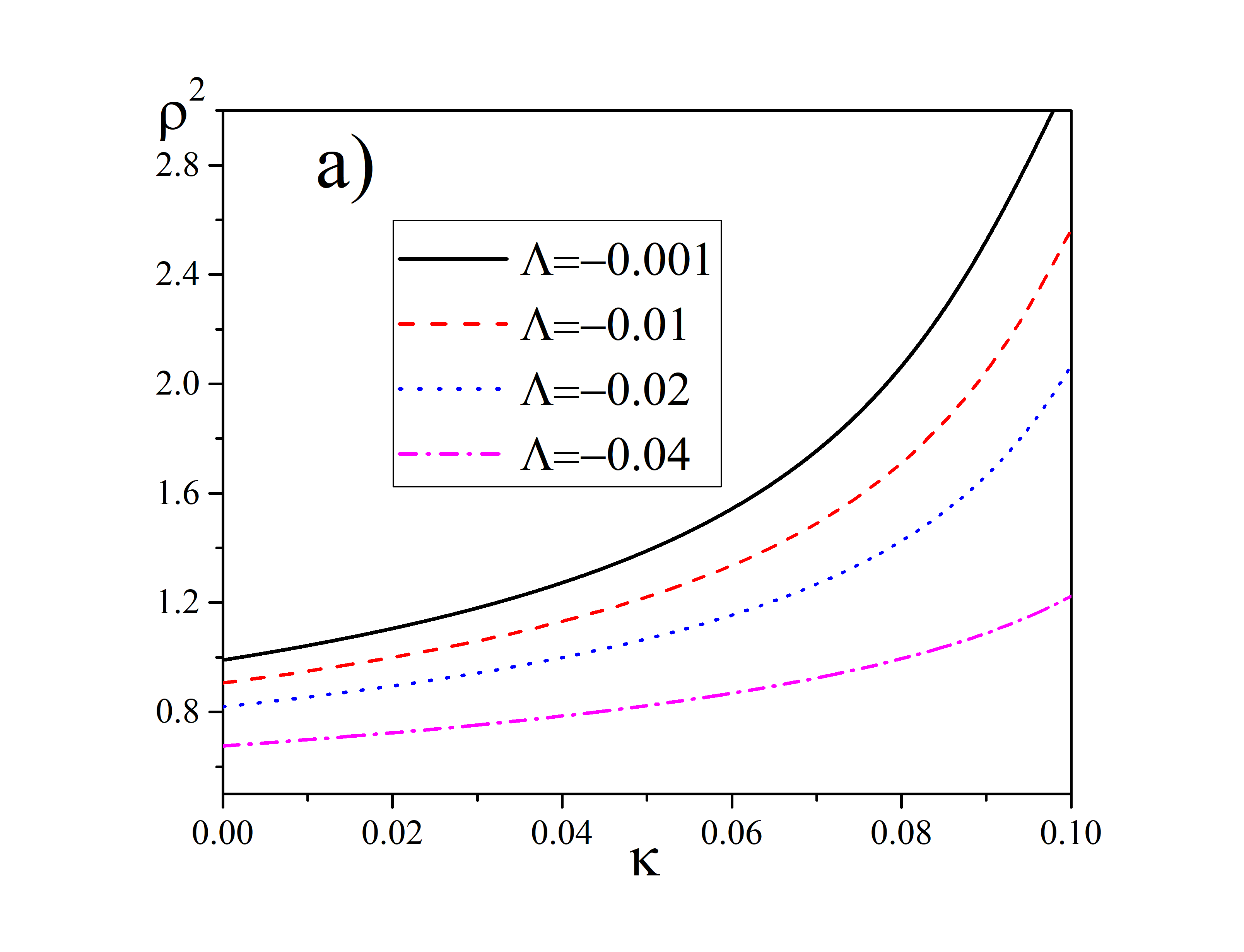}
  \end{minipage} \quad
  \begin{minipage}{.48\textwidth}
  \includegraphics[width=8cm]{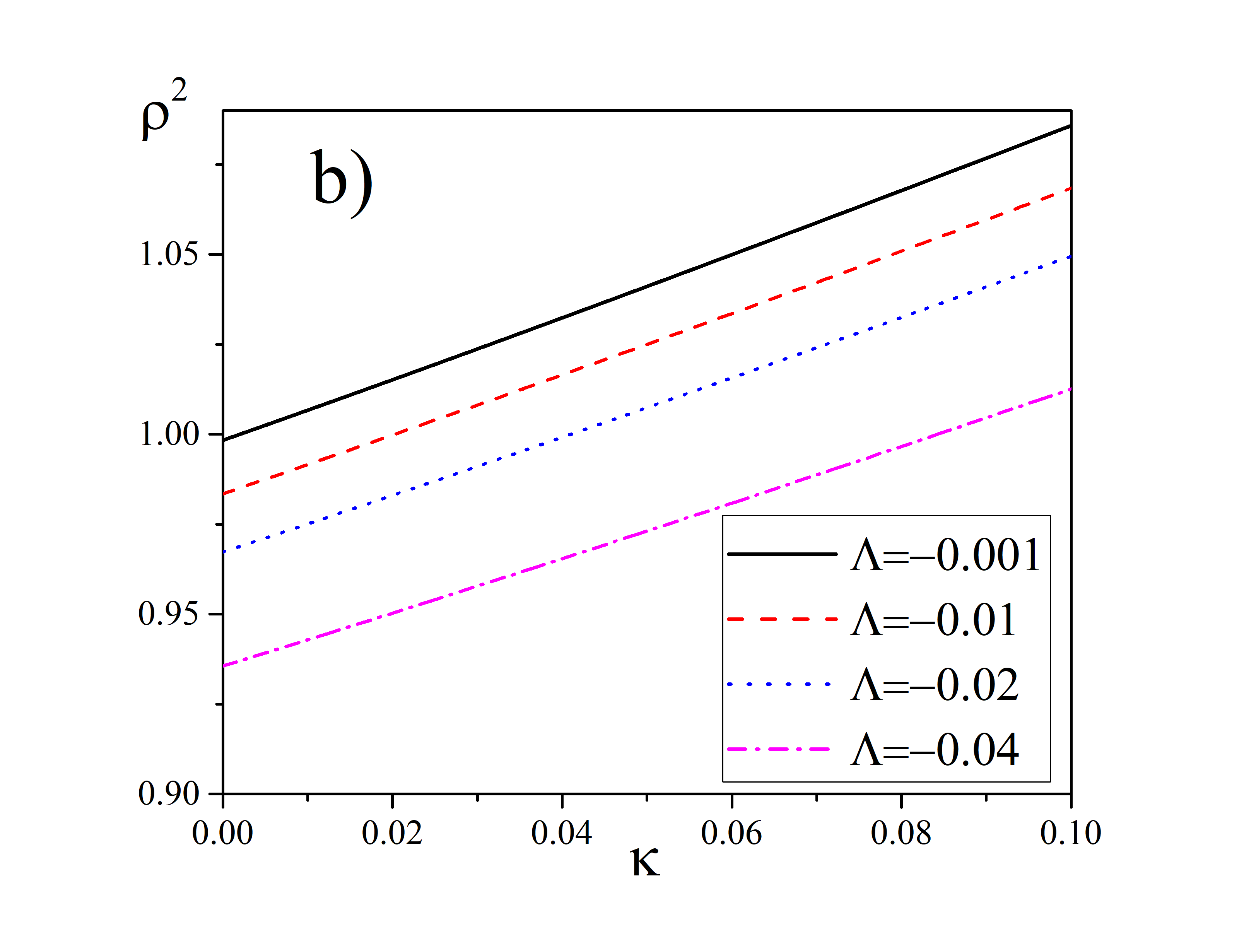}
\end{minipage} \quad
\caption{\label{fig10} Dependence of the beam area $\rho^2$ on the graded dissipation parameter $\kappa$ for different saturated net-loss parameters $\Lambda$. $\tau=0.1$ (a) and 1 (b); $\delta=-1$, $s=-1$.}
\end{figure*}

The corresponding fourth-order polynomial equation for $\rho^2$, which closes the system, is too cumbersome to be written out explicitly here (see \emph{Mathematica} notebook \cite{nb}). Only one of the solutions of this equation has a physical meaning: examples of this solution are shown in Fig. \ref{fig10}, for different values of the dissipation parameter $\Lambda$ and $\kappa$, and the normal GVD regime $\delta=-1$.

In the case of the DS described by Eq. (18), a decrease of the effective aperture $\chi$ results in beam squeezing, accompanied by a growth of the peak power (see Figs. \ref{fig6}, \ref{fig7}). Whereas, for the DS expressed by Eqs. (24--27), decreasing $\chi$ leads to a widening of the beam (see Fig. \ref{fig10}, as well as Fig. 2 in the main text). This means that different mechanisms of DS formation are in place for the two cases. If in the former case, spatial and temporal mechanisms are only related through the DS amplitude, which is defined by the effective beam size. On the other hand, in the latter case one deals with an interplay between nondissipative and dissipative mechanisms, as it occurs in mode-locked lasers operating in either the anomalous or the normal GVD regime. This conclusion is supported by the fact that the DS described by Eqs. (24--27) also exists in the nondissipative limit. Dissipative factors may stabilize a soliton, however their overdoing leads to soliton degradation.

The impact of dynamical gain into the $\Lambda$-parameter needs some clarification (see Eq. (15)). A comparison between Fig. \ref{fig10} and Fig. \ref{fig2} from the main text reveals that a perturbation leading to a growth of the DS energy (and thereby, reducing $|\Lambda|$) would also increase the DS temporal duration and beam area, and decrease its peak peak power (and vice versa). That could prevent both DS collapse and degradation due to spatial spreading. Thus, the overall effect of dynamical gain saturation would play the role of a negative passive feedback, and provide DS robustness. However, the validity of this conjecture needs further study.

\subsection{\label{sec:num}The numerical study of the dissipative Gross-Pitaevskii equation}

Numerical simulations were based on two approaches: i) solution of the ordinary differential equations (17) in the framework of the VA, and ii) direct numerical solution of Eq. (14). In the first case, the primary intention was to inspect the ``attraction basin'' of the solution (24--27). Whereas, in the second case, we aimed at investigating the propagation regimes extending beyond the lowest-mode soliton-like approximation of Eq. (9).

\begin{figure*}[b]
  \begin{minipage}{.48\textwidth}
\includegraphics[width=8cm]{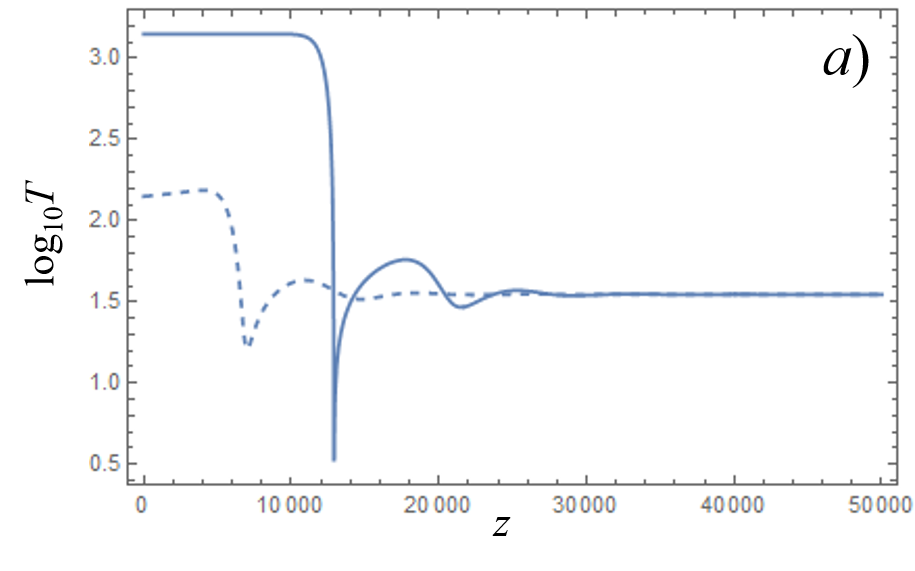}
  \end{minipage} \quad
  \begin{minipage}{.48\textwidth}
  \includegraphics[width=8cm]{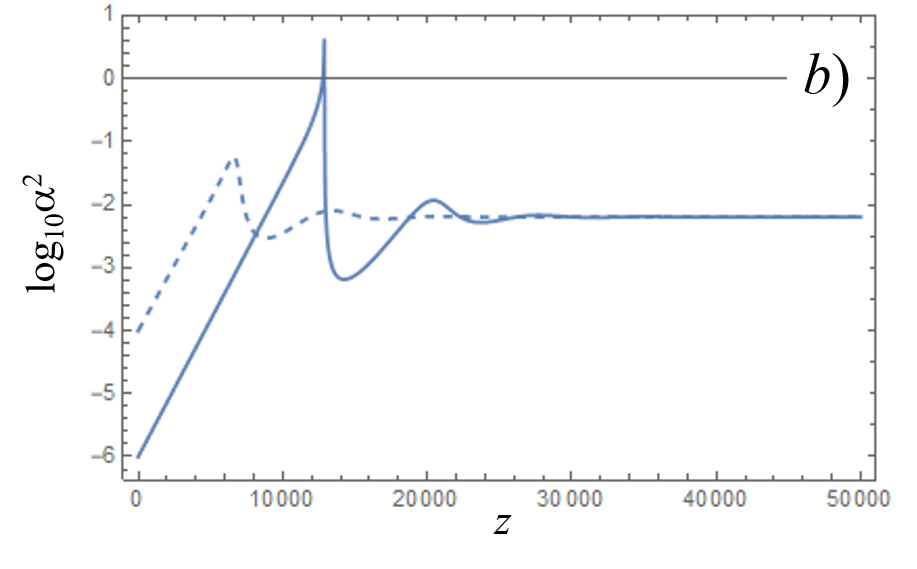}
\end{minipage} \quad
\caption{\label{fig11} Evolution of the DS duration $T$ (\emph{a}) and its intensity $\alpha^2$ (\emph{b}) from the system (17) for the initial ``seed'' $\alpha(0) =10^{-3}$ (solid curve) and $\alpha(0) =10^{-2}$ (dashed curve). $\tau=0.5$, $\Lambda=-0.0015$, $\kappa=0.001$, $\delta=-1$, $s=-1$.}
\end{figure*}

Fig. \ref{fig11} demonstrates the evolution of the DS temporal duration and peak power for two different ``seed'' amplitudes $\alpha(0)$ in the normal GVD regime with other initial conditions corresponding to Eq. (5): $T(0) = {{\sqrt {{{2\left| \delta  \right|} \mathord{\left/
 {\vphantom {{2\left| \delta  \right|} \nu }} \right.
 \kern-\nulldelimiterspace} \nu }} } \mathord{\left/
 {\vphantom {{\sqrt {{{2\left| \delta  \right|} \mathord{\left/
 {\vphantom {{2\left| \delta  \right|} \nu }} \right.
 \kern-\nulldelimiterspace} \nu }} } {\alpha \left( 0 \right)}}} \right.
 \kern-\nulldelimiterspace} {\alpha \left( 0 \right)}},\,\rho \left( 0 \right) = {{\left( {\nu \,\alpha {{\left( 0 \right)}^2} - \sqrt {{\nu ^2}\alpha {{\left( 0 \right)}^4} - 36s} } \right)} \mathord{\left/
 {\vphantom {{\left( {\nu \,\alpha {{\left( 0 \right)}^2} - \sqrt {{\nu ^2}\alpha {{\left( 0 \right)}^4} - 36s} } \right)} {6s}}} \right.
 \kern-\nulldelimiterspace} {6s}},\,\,\psi \left( 0 \right) = 0,\,\theta \left( 0 \right) = 0$. The existence of such convergent solutions underlies the stability borders marked by the scatter symbols in Fig. 3 of the main text.

One can see that the DS ``attracting basin'' is broad. Also, the analysis demonstrates that such a ``basin'' broadens with the growth of $\tau$, which is a spectral dissipation enhancement. Nevertheless, Fig. \ref{fig11} shows a ``blow-up'' dynamics of the growing DS, that can prevent its formation.

As the VA is constrained by a pre-defined ansatz, direct numerical simulations are required. We performed simulations of Eq. (14) by using the finite-element method implemented by COMSOL Multiphysics software. These simulations reveal the existence of new scenarios, whose existence is beyond the reach of the VA approximation. Specifically: i) multimode dynamics in the normal GVD regime and ii) multipulsing in the anomalous GVD regime.

The video on \href{http://info.tuwien.ac.at/kalashnikov/multimode.avi}{multimode evolution} \cite{mm} corresponding to the inset in Fig. 3, \emph{b} of the main text illustrates the first scenario (vertical axis $t\in\left\{-60,60\right\}$, horizontal axis $r\in\left\{0,7\right\}$, propagation interval $z\in\left\{0,1500\right\}$, $\delta=-1$, $\tau=1$, $\chi=4$, $\kappa=0.001$, $\alpha=0.1$). One can see the presence of non-trivial spatiotemporal dynamics, which affects spatiotemporal mode-locking.

The video on \href{http://info.tuwien.ac.at/kalashnikov/multipulse.avi}{multipulse evolution} \cite{mp}, corresponding to inset in Fig. 3, \emph{a} of the main text, displays the spatiotemporal multipulse dynamics (vertical axis $t\in\left\{-60,60\right\}$, horizontal axis $r\in\left\{0,7\right\}$, propagation interval $z\in\left\{0,1500\right\}$, $\delta=-1$, $\tau=1$, $\chi=4$, $\kappa=0.001$, $\alpha=0.1$). In this case, the temporal dynamics prevails on the spatial one, so that an initial low-intensity DS splits into several interacting pulses.

As an outlook for further work, we would like to mention the study of the DS dynamics under the influence of dynamical gain saturation, and the impact of the multimodal field structure in a MMF. In addition, more complex transverse waveguide structures, such as a PCF, could be investigated. One may hope that such structures could provide additional mechanisms for the self-starting of the DS, its stabilization, and for decreasing the DS peak power, all of which is desirable for implementing the concept of DKLM in a practical guided wave laser architecture.

\bibliographystyle{unsrt}
\bibliography{references}  

\begin{thebibliography}{10}

\bibitem{Malomed1}
B.~A. Malomed.
\newblock {\em Eur. Phys. J. Special Topics}, 225:2507--2532, 2016.

\bibitem{Malomed2}
Y.~V. Kartashov, G.~A. Astrakharchik, B.~A. Malomed, and L.~Torner.
\newblock {\em Nature Reviews}, 1:185--197, 2019.

\bibitem{Wabnitz1}
K.~Krupa, A.~Tonello, A.~Barth\'{e}l\'{e}my, T.~Mansuryan, G.~Millot
  V.~Couderc, Ph. Grelu, D.~Modotto, S.~A. Babin, and S.~Wabnitz.
\newblock {\em APL Photonics}, 4:110901, 2019.

\bibitem{Serkin}
V.~N. Serkin and T.~L. Belyaeva.
\newblock {\em Optik}, 176:38--48, 2019.

\bibitem{Desaix}
M.~Karlsson, D.~Anderson, and M.~Desaix.
\newblock 1992.

\bibitem{Wang}
Sh.-Sh. Yu, Ch.-H. Chien, Y.~Lai, and J.~Wang.
\newblock {\em Optics Commun.}, 119:167--170, 1995.

\bibitem{Agrawal1}
S.~Raghavan and G.~P. Agrawal.
\newblock {\em Optics Commun.}, 180:377--382, 2000.

\bibitem{Akhmediev}
Ph. Grelu, J.~M. Soto-Crespo, and N.~Akhmediev.
\newblock {\em Optics Express}, 13:9352--9360, 2005.

\bibitem{Malomed3}
Thawatchai Mayteevarunyoo, Boris~A. Malomed, and Dmitry~V. Skryabin.
\newblock {\em Opt. Express}, 27, 2019.

\bibitem{np}
Editorial.
\newblock {\em Nature Photonics}, 8:1, 2014.

\bibitem{Wise1}
W.~H. Renninger and F.~W. Wise.
\newblock {\em Nature Commun.}, 4:1719, 2013.

\bibitem{Wabnitz2}
R.~Guenard, K.~Krupa, R.~Dupiol, M.~Fabert, A.~Bendahmane, V.~Kermene,
  A.~Desfarges-Berthelemot, J.~L. Auguste, A.~Tonello, A.~Barth\'{e}l\'{e}my,
  G.~Millot, S.~Wabnitz, and V.~Couderc.
\newblock {\em Optics Express}, 25:22219, 2017.

\bibitem{Agrawal2}
A.~S. Ahsan and G.~P. Agrawal.
\newblock {\em Optics Lett.}, 43:3345, 2018.

\bibitem{Beach}
T.~Bhutta, J.~I. Mackenzie, D.~P. Shepherd, and R.~J. Beach.
\newblock {\em J. Opt. Soc. Am. B}, 19:1539--1543, 2002.

\bibitem{Wise2}
L.~G. Wright, P.~Sidorenko, H.~Pourbeyram, Z.~M. Ziegler, A.~Isichenko, B.~A.
  Malomed, C.~R. Menyuk, D.~N. Christodoulides, and F.~W. Wise.
\newblock 2020.

\bibitem{Moser}
U.~Tegin, E.~Kakkava, B.~Rahmani, D.~Psaltis, and Ch. Moser.
\newblock {\em Optica}, 6:1412--1415, 2019.

\bibitem{Akhmediev2}
Ph. Grelu and N.~Akhmediev.
\newblock {\em Nature Photonics}, 6:84--92, 2012.

\bibitem{Fermann}
M.~E. Fermann, A.~Galvanauskas, and G.~Sucha.
\newblock {\em Ultrafast Lasers: Technology and Applications}.
\newblock Marcel Dekker, 2003.

\bibitem{Mackenzie}
J.~I. Mackenzie.
\newblock {\em IEEE Journal of Selected Topics in Quantum Electronics},
  13:626--637, 2007.

\bibitem{Kartner}
H.~{Byun}, D.~{Pudo}, S.~{Frolov}, A.~{Hanjani}, J.~{Shmulovich}, E.~P.
  {Ippen}, and F.~X. {Kartner}.
\newblock {\em IEEE Photonics Technology Letters}, 21:763--765, 2009.

\bibitem{Sibbett}
D.~E. Spence, P.~N. Kean, and W.~Sibbett.
\newblock {\em Opt. Lett}, 16:42--44, 1991.

\bibitem{Pronin}
J.~Zhang, J.~Brons, M.~Seidel, D.~Bauer, D.~Sutter, V.~Pervak, V.~Kalashnikov,
  Z.~Wei, A.~Apolonski, F.~Krausz, and O.~Pronin.
\newblock {\em Advanced Solid State Lasers Conference, OSA Technical Digest},
  page ATh4A.7, 2015.

\bibitem{Wise3}
W.~H. Renninger and F.~W. Wise.
\newblock {\em Optica}, 1:101--104, 2014.

\bibitem{Wise4}
L.~G. Wright, W.~H. Renninger, D.~N. Christodoulides, and F.~W. Wise.
\newblock {\em Optics Express}, 23, 2015.

\bibitem{Wise5}
L.~G. Wright, Z.~M. Ziegler, P.~M. Lushnikov, Z.~Zhu, M.~A. Eftekhar, D.~N.
  Christodoulides, and F.~W. Wise.
\newblock {\em IEEE J. Selected Topics in Quantum Electron.}, 24:5100516, 2018.

\bibitem{Pitaevskii}
L.~Pitaenskii and S.~Stringari.
\newblock {\em Bose-Einstein Condensation and Superfluidity}.
\newblock Oxford Univ. Press, 2016.

\bibitem{Kant}
S.~Ch. Cerda, S.~B. Cavalcanti, and J.~M. Hickmann.
\newblock {\em Eur. Phys. J. D.}, 1:313--316, 1998.

\bibitem{Janner}
S.~Longhi and D.~Janner.
\newblock {\em J. Opt. B: Quantum Semiclass. Opt.}, 6:S303--S308, 2004.

\bibitem{Pronin2}
O.~Pronin, J.~Brons, C.~Grasse, V.~Pervak, G.~Boehm, M.-C. Amann, A.~Apolonski,
  V.~L. Kalashnikov, and F.~Krausz.
\newblock {\em Optics Lett.}, 37:3543--3545, 2012.

\bibitem{Wise6}
L.~G. Wright, D.~N. Christodoulides, and F.~W. Wise.
\newblock {\em Science}, 358:94--97, 2017.

\bibitem{Paul}
V.~L. Kalashnikov and S.~V. Sergeyev.
\newblock In C.~Paul, editor, {\em Fiber Laser}, pages 165--210. InTechOpen,
  2016.

\bibitem{negative}
M.~V. Berry.
\newblock {\em J. Phys. A: Math. Theor.}, 43:415302, 2010.

\bibitem{talbot}
J.~Wen and Y.~Zhang.
\newblock {\em Advances in Optics and Photonics}, 5:83--130, 2013.

\bibitem{muller}
D.~E. Muller.
\newblock {\em Mathematical Tables and Other Aids to Computation}, 10:208--215,
  1956.

\bibitem{nb}
V.~L. Kalashnikov.
\newblock {\em Variational approach to a fiber-laser spatial-temporal
  dissipative soliton (Mathematica notebook)}, 2020.
\newblock \url{http://info.tuwien.ac.at/kalashnikov/Variational_Gaussian.nb}.

\bibitem{mm}
V.~L. Kalashnikov.
\newblock {\em Multimode evolution of dissipative soliton (video)}, 2020.
\newblock \url{http://info.tuwien.ac.at/kalashnikov/multimode.avi}.

\bibitem{mp}
V.~L. Kalashnikov.
\newblock {\em Multimode evolution of multiple dissipative solitons (video)},
  2020.
\newblock \url{http://info.tuwien.ac.at/kalashnikov/multipulse.avi}.

\end{thebibliography}






\end{document}